\documentclass{French}%{parris}
\usepackage{latexsym}
\usepackage[french]{babel}
\usepackage{psfrag}
\usepackage{amsmath}
\usepackage{amssymb}

\usepackage{color}

\begin{document}
\subh{Improving resource elasticity in cloud computing thanks to model-free control}
\author{Maria Bekcheva\ts{1}, Michel Fliess\ts{2,4}, C\'{e}dric Join\ts{3,4}, Alireza Moradi\ts{5}, Hugues Mounier\ts{1}}
\aff{\removelastskip \ts{1}Laboratoire des Signaux et Syst\`emes (L2S),  Universit\'e Paris-Sud-CNRS-CentraleSup\'elec, Universit\'e Paris-Saclay, 
91192 Gif-sur-Yvette, France, \{maria.bekcheva, hugues.mounier\}@l2s.centralesupelec.fr\\
\ts{2}LIX (CNRS, UMR 7161), \'Ecole polytechnique, 91128 Palaiseau, France, Michel.Fliess@polytechnique.edu\\
\ts{3}CRAN (CNRS, UMR 7039), Universit\'{e} de Lorraine, BP 239, 54506 Vand{\oe}uvre-l\`{e}s-Nancy, France, cedric.join@univ-lorraine.fr\\
\ts{4}AL.I.E.N. (ALg\`{e}bre pour Identification \& Estimation Num\'{e}riques), 7 rue Maurice Barr\`{e}s, 54330 V\'{e}zelise, France, \{michel.fliess, cedric.join\}@alien-sas.com \\
\ts{5} Inagral, 128 rue  de la Bo\'{e}tie, 75008 Paris, France, alireza@inagral.com }

\resu{L'adaptation dynamique des ressources de calcul \`a des variations de trafic, dans la gestion \og{nuagique}\fg, est un domaine actif d'investigation. Les automaticiens ont d\'{e}j\`a propos\'{e} maints rem\`{e}des. On emploie, ici, la commande sans mod\`{e}le et les correcteurs \og{intelligents}\fg \ associ\'{e}s, faciles \`a implanter et aux nombreux succ\`{e}s industriels, pour traiter l'\og{\'{e}lasticit\'{e} horizontale}\fg. Le comportement, compar\'{e} aux algorithmes commerciaux d'auto-ajustement, est meilleur, m\^{e}me avec des fluctuations aig\"{u}es de charge. Des exp\'{e}riences sur le service Web d'Amazon (AWS) le confirment.} 
\abstract{In cloud computing management, the dynamic adaptation of computing resource allocations under time-varying workload is an active domain of investigation. Several control strategies were already proposed. Here the model-free control setting and the corresponding ``intelligent'' controllers, which are most successful in many concrete engineering situations, are employed for the ``horizontal elasticity.'' When compared to the commercial ``Auto-Scaling'' algorithms, our easily implementable approach, behaves better even with sharp workload fluctuations. This is confirmed by experiments on Amazon Web Services (AWS).}
\moi{\textbf{\textcolor{AbsBlue}{MOTS-CL\'{E}S.}}\hphantom{--} Nuagique, allocation des ressources de calcul, ajustement, \'{e}lasticit\'{e}, commande sans mod\`{e}le, AWS}
\kwd{Cloud computing, computing resources allocation, scaling, elasticity, model-free control, AWS}

\chapter{Meilleure \'{e}lasticit\'{e} \og{nuagique}\fg \ par commande sans mod\`{e}le}
\begin{flushright}
\textit{This is not the end. It is not even the beginning of the end. \\ But it is, perhaps, the end of the beginning.} \\
Churchill (10 novembre 1942)
\end{flushright}

\section{Introduction \vspace*{-5truemm}}
\subsection{Prol\'{e}gom\`{e}nes}

Nul besoin d'acqu\'{e}rir et de ma\^{\i}triser mat\'{e}riels et logiciels chers. 
L'essentiel se fait ailleurs, via Internet. C'est le but du \og nuagique \fg, raccourci 
d'\og infonuagique \fg, invent\'{e} au Qu\'{e}bec pour traduire \emph{cloud 
computing}\footnote{La presse fran\c{c}aise \'{e}crit souvent \og informatique 
d\'{e}mat\'{e}rialis\'{e}e \fg. C'est long et, surtout, trompeur. Un mat\'{e}riel lourd 
est toujours l\`a, mais loin de l'utilisateur.}. 
Voici quelques \'{e}chantillons \cite{berkeley,buy}, \\ \cite{fehling,marinescu,wu} de la vaste 
litt\'{e}rature consacr\'{e}e \`a cette technologie informatique en croissance rapide. 
Un march\'{e} consid\'{e}rable se d\'{e}veloppe. Amazon, Google, Microsoft, IBM, Oracle aux 
\'Etats-Unis, Alibaba, China Telecom, Tencent en Chine, NTT Communications au Japon, SAP en Allemagne, Orange, OVH en France sont parmi les entreprises les plus connues. La facult\'{e} d'ajuster au mieux et en temps r\'{e}el les ressources de calcul réclam\'{e}es par l'usager, qui est un avantage cl\'{e} du nuagique, s'appelle \og{\'{e}lasticit\'{e}}\fg \ (voir, par exemple, \cite{herbst}). Ce sujet tr\`{e}s actif de recherches est, ici, le n\^{o}tre.
%On propose, ici, une solution par 
%\emph{commande sans mod\`{e}le} \cite{ijc13}, aussi nouvelle qu'efficace, pour une 
%meilleure allocation des ressources de calculs, question cl\'{e} s'il en est. 

Avant de passer au vif, quelques remarques s'imposent:
\begin{itemize}
\item \'Ecrire en fran\c{c}ais est une gageure. La domination absolue de l'anglais lui a donn\'{e} un statut subalterne, notamment dans le champ scientifique. S'ajoute, dans une discipline neuve, la difficult\'{e}, consid\'{e}rable, de trouver des \'{e}quivalents du vocabulaire am\'{e}ricain\footnote{Les publications en fran\c{c}ais sont rares (voir, par exemple, \cite{rivard,vicat}). La situation des autres \og{grandes}\fg \ langues scientifiques occidentales, comme l'allemand, est similaire.}. Que l'impact de cette contribution n'en soit point trop amoindri!
\item Kalman (voir \cite{kalman} et, aussi, \cite{eilenberg,sontag}) avait tent\'{e} d'\'{e}tablir un lien entre automatique, c'est-\`a-dire, alors, syst\`{e}mes lin\'{e}aires de dimension finie, et th\'{e}orie des automates finis, adonc dominante en informatique th\'{e}orique. Le lien, vieux de pr\`{e}s de soixante ans \cite{schutz} entre automates finis, mono\"{\i}des libres et s\'{e}ries rationnelles non commutatives (voir, aussi, \cite{christophe,saka}) rattache, en fait, ces automates 
aux syst\`{e}mes \og{bilin\'{e}aires}\fg, ou \og{r\'{e}guliers}\fg, de l'automatique  gr\^{a}ce aux s\'{e}ries g\'{e}n\'{e}ratrices non commutatives rationnelles \cite{informcontr}. Cette probl\'{e}matique, aussi valable soit-elle, a vieilli. L'interaction entre informatique et automatique a migr\'{e} vers des sujets plus concrets (voir, par exemple, \cite{hellerstein,janert,leva,marinescu}). L'investigation des rapports entre automatique et informatique demeure n\'{e}anmoins mineure \`a l'universit\'{e} et dans l'industrie. 
\item Voici un rajout quelque peu sarcastique. L'intelligence artificielle, qu'on ne sait d\'{e}finir vraiment, est vue aujourd'hui comme l'acm\'{e} de l'informatique, non seulement dans les m\'{e}dias, mais aussi par moult acteurs politiques, 
\'{e}conomiques, administratifs et universitaires. D'aucuns (voir, par exemple, \cite{chong} et sa bibliographie), dans des cercles d'ingénieurs, l'appréhendent comme un avatar de l'automatique.
\end{itemize}

\subsection{Rudiments d'\'{e}lasticit\'{e}}
Afin de r\'{e}pondre \`a des variations, parfois brutales, de charges, l'approche traditionnelle provisionne 
\`a l'avance, au risque d'un fort gaspillage, des ressources consid\'{e}rables. En nuagique, la puissance de calculs est li\'{e}e aux \og machines virtuelles \fg, ou \emph{virtual machines} (\emph{VM}), c'est-\`a-dire \`a des \'{e}mulations d'ordinateurs dues \`a des logiciels. Elles sont h\'{e}berg\'{e}es sur des serveurs physiques mutualis\'{e}s. Un regroupement de machines virtuelles est appel\'{e} \og grappe \fg, ou \emph{cluster}. L'\og{\'{e}lasticit\'{e} horizontale}\fg \ modifie la cardinalit\'{e} de la grappe sans en modifier les éléments\footnote{Quant \`a l'\og{\'{e}lasticit\'{e} verticale}\fg, elle alt\`{e}re la structure des machines virtuelles et/ou physiques}. 

Comme en t\'{e}moignent les synth\`{e}ses dues \`a \cite{al,Galante2012}, \\ \cite{lorido,pati,ullah}, de multiples \\ techniques de commande ont \'{e}t\'{e} propos\'{e}es pour assurer un meilleur \og{ajustement}\fg, ou \emph{scaling}. Les PID, primordiaux dans l'industrie classique (voir, par exemple, \cite{astrom}, \\ \cite{murray,od}) y dominent aussi.  L'explication reste identique: impossibilit\'{e} d'une description math\'{e}matique exploitable pour la plupart des applications concr\`{e}tes, quel qu'en soit le domaine.

\subsection{Notre approche} 
On utilise la \og{commande sans mod\`{e}le}\fg, ou \emph{model-free control} (\emph{MFC}), \cite{ijc13}. Elle
\begin{itemize}
\item conserve les avantages des PID  sans en avoir les graves inconv\'{e}nients \cite{ijc13}, \\ \cite{iste},
\item a connu nombre de succ\`{e}s concrets (voir une liste assez compl\`{e}te, du moins jusqu'au d\'{e}but 2018, dans les bibliographies de \cite{ijc13,bara}),
\item est facile \`a implanter \cite{ijc13,nice}.
\end{itemize}
\begin{remark}
On trouve en \cite{ijc13} des r\'{e}f\'{e}rences sur l'emploi du vocable \newline \og{sans-mod\`{e}le}\fg \ en automatique, mais avec des sens tout diff\'{e}rents. Il en va de m\^{e}me en nuagique (voir, par exemple, \cite{bu,rao,wang}).
\end{remark}

L'algorithme est  test\' e  \`{a} l'aide d'\emph{Amazon Web Services}; l'acronyme \emph{AWS} est connu de la plupart\footnote{ Renvoyons, pour toute précision utile ici, à \cite{wittig}) et, surtout, au lien, fourni par Amazon, \\
  {\tt https$:$//docs.aws.amazon.com/fr\_fr/AWSEC2/latest/UserGuide/concepts.html} \\ Il importe de souligner que l'un des auteurs, A. Moradi, est \emph{AWS Certified Solutions Architect - Associate}.}. Les comparaisons avec deux d\'{e}marches:
\begin{itemize}
\item sans \og{auto-ajustement}\fg, ou \emph{auto-scaling},
\item et, surtout, avec l'\og{ajustement pour suivi de cible}\fg, ou \emph{target tracking scaling}, d'Amazon \emph{Elastic Compute Cloud} (\emph{EC2}), %(voir, par exemple, \cite{vlet} et \cite{amazonGuide}),
\end{itemize} 
tournent largement en faveur du sans-mod\`{e}le (voir tableau {\ref{Tab_1}}).

%\textcolor{green}{Rajouter un r\'{e}sum\'{e} des r\'{e}sultats quantitatifs obtenus et de leurs comparaisons avec d'autres.}

{\subsection{Plan}}

On trouve des rappels sur cette commande au paragraphe {\ref{rappel}}. Les paragraphes {\ref{mise}} et {\ref{exp}} d\'{e}taillent respectivement mise en {\oe}uvre, exp\'{e}rimentations et comparaisons. La conclusion au paragraphe {\ref{conclusion} }sugg\`{e}re non seulement  de futures pistes mais aussi des r\'{e}flexions quant \`{a} la recherche nuagique.

\newpage

\section{Br\`{e}ve \'{e}vocation du sans-mod\`{e}le{\protect\footnote{Renvoyons \`a \cite{ijc13} pour plus de d\'{e}tails.}}}
\label{rappel}
\subsection{Mod\`{e}le ultra-local et correcteur intelligent}
Soit un syst\`{e}me entr\'{e}e-sortie monovariable, c'est-\`a-dire avec une seule commande $u$ et une seule sortie $y$, dont la description math\'{e}matique est inextricable. Ce fait conduit \`a introduire le mod\`{e}le \og{ultra-local}\fg:
\begin{equation*}
y^{(\nu)} = F + \alpha u \label{0}
\end{equation*}
o\`{u} $\nu \geq 1$.
\begin{itemize}
\item En g\'{e}n\'{e}ral, $\nu = 1$: 
\begin{equation}
\boxed{\dot{y} = F + \alpha u} \label{ul}
\end{equation}
\item Le praticien choisit le param\`{e}tre constant $\alpha \in \mathbb{R}$  de sorte que les trois termes de \eqref{ul} soient de magnitudes semblables. Une identification pr\'{e}cise de $\alpha$ est sans objet.
\item $F$, qui confond structure inconnue du syst\`{e}me et perturbations externes, s'estime \`a chaque instant \`a partir de $u$ et $y$.
\end{itemize}

 On associe \`a \eqref{ul}  le correcteur \og{intelligent proportionnel}\fg, ou \emph{iP}, 
\begin{equation}\label{ip}
\boxed{u = - \frac{F_{\rm estim} - \dot{y}_d - K_P e}{\alpha}}
\end{equation}
\begin{itemize}
\item $F_{\rm estim}$ est une estim\'{e}e $F$.
\item $y_d$ est la trajectoire de r\'{e}f\'{e}rence, ou consigne,
\item $e = y_d - y$ est l'erreur de poursuite,
\item $K_P \in \mathbb{R}$ est un gain.
\end{itemize}
Il vient, d'apr\`{e}s \eqref{ul} et \eqref{ip},
\begin{equation*}\label{track}
\dot{e} + K_P e = F_{\rm estim} - F
\end{equation*}
Le choix d'un $K_P$ stabilisant est transparent. La poursuite est \og{bonne}\fg \ si l'estim\'{e}e $F_{\rm estim}$ l'est, c'est-\`a-dire si $F - F_{\rm estim} \simeq  0$.

\subsection{Estimation de $F$}\label{F}
\subsubsection{Premi\`{e}re formule}
D'apr\`{e}s une propri\'{e}t\'{e} classique d'analyse mathématique (voir, par exemple, \cite{godement}), on peut, sous des hypoth\`{e}ses faibles, approcher $F$ en \eqref{ul} par une fonction $F_{\text{estim}}$, constante par morceaux. Avec les notations du calcul op\'{e}rationnel  (voir, par exemple, \cite{erde}), \eqref{ul}  s'\'{e}crit:
$$
sY = \frac{\Phi}{s}+\alpha U +y(0)
$$
o\`{u} $\Phi$ est une constante. On \'{e}limine la condition initiale $y(0)$ en d\'{e}rivant les deux membres par $\frac{d}{ds}$:
$$
Y + s\frac{dY}{ds}=-\frac{\Phi}{s^2}+\alpha \frac{dU}{ds}
$$
On multiplie \`a gauche les deux membres par $s^{-2}$. D'o\`{u}, dans le domaine temporel, une estim\'{e}e en temps r\'{e}el, obtenue gr\^{a}ce \`a l'\'{e}quivalence entre $\frac{d}{ds}$ et la multiplication par $-t$,
\begin{equation}\label{integral1}
{\small F_{\text{estim}}(t)  =-\frac{6}{\tau^3}\int_{t-\tau}^t \left\lbrack (\tau -2\sigma)y(\sigma)+\alpha\sigma(\tau -\sigma)u(\sigma) \right\rbrack d\sigma}
\end{equation}

\subsubsection{Seconde formule}
\label{2e}
En utilisant la définition \eqref{ip} de l'iP, il vient:
\begin{equation}\label{integral2}
F_{\text{estim}}(t) = \frac{1}{\tau}\left[\int_{t - \tau}^{t}\left(\dot{y}_{d}-\alpha u
- K_P e \right) d\sigma \right] 
\end{equation}
\begin{remark}
Quelques points utiles:
\begin{itemize}
\item Les calculs \eqref{integral1} et \eqref{integral2} se font en temps r\'{e}el. 
\item Les int\'{e}grales en \eqref{integral1} et \eqref{integral2} sont des filtres passe-bas, qui att\'{e}nuent le bruit. 
\item D'un point de vue pratique,  l'échantillonnage m\`{e}ne \`a des filtres num\'{e}riques, faciles \`a implanter.
\end{itemize}
\end{remark}

\section{Mise en {\oe}uvre}\label{mise}
On utilise le mod\`{e}le ultra-local \eqref{ul} et le correcteur iP \eqref{ip}. 
\begin{itemize}
\item La sortie $y$ correspond à l'utilisation des \og processeurs \fg, ou \emph{central processing units} (\emph{CPU}), durant le p\'{e}riode $[t-h, t)$ d'\'{e}chantillonnage:
\begin{itemize}
\item Soit $\text{CPU}^h_{i}(t)$ la charge moyenne, sur l'intervalle $[t-h, t)$, du processeur de la machine virtuelle d'ordre $i$, d\'{e}pendant des requ\^{e}tes. 
\item Soit $M_{\text{act}}^h(t)$, $M_{\rm{min}} \leqslant M_{\text{act}}^h(t) \leqslant M_{\text{max}}$, le nombre de machines virtuelles actives.
\end{itemize}
\noindent Alors 
\begin{equation} \label{Ymesured}
y(t) = \sum_{i = 1}^{ M_{\rm{act}}^h(t)} \text{CPU}^h_{i}(t)
\end{equation}
\item Pour la trajectoire d\'{e}sir\'{e}e, ou consigne, $y_d$, on choisit 
\begin{equation}\label{Ydesired}
y_d(t) = \frac{M_{\rm{act}}^h(t)}{2}
\end{equation}
Ce choix est un compromis. Ainsi, une consigne 
\'{e}gale \`a $0.3 \times M_{\rm{act}}^h(t)$ (resp. $0.8 \times M_{\rm{act}}^h(t)$) impliquerait une sous-exploitation (resp. sur-exploitation). Ajoutons que toute sur-exploitation notable induit un retard significatif dans l'exécution des requ\^{e}tes. D'o\`{u} une d\'{e}gradation de la qualit\'{e} de service.
\item La commande $u$ correspond \`a la cardinalit\'{e} de la grappe, c'est-\`a-dire au nombre de machines virtuelles actives.
\end{itemize}
%\begin{remark}
%Certaines valeurs en ordonnées dans les figures du paragraphe {\ref{pourc}} sont données en pourcentages. C'est pourquoi $1/2$ correspond ainsi à $50$. 
%\end{remark}

\begin{remark}
Certaines valeurs en ordonnées dans les figures du paragraphe {\ref{pourc}} sont données en pourcentages de façon évidente. C'est pourquoi, par exemple, $1/2$ en \eqref{Ydesired} correspond alors à $50\%$. 
\end{remark}

\begin{figure*}[h!]
\begin{center}
\includegraphics[width=7in]{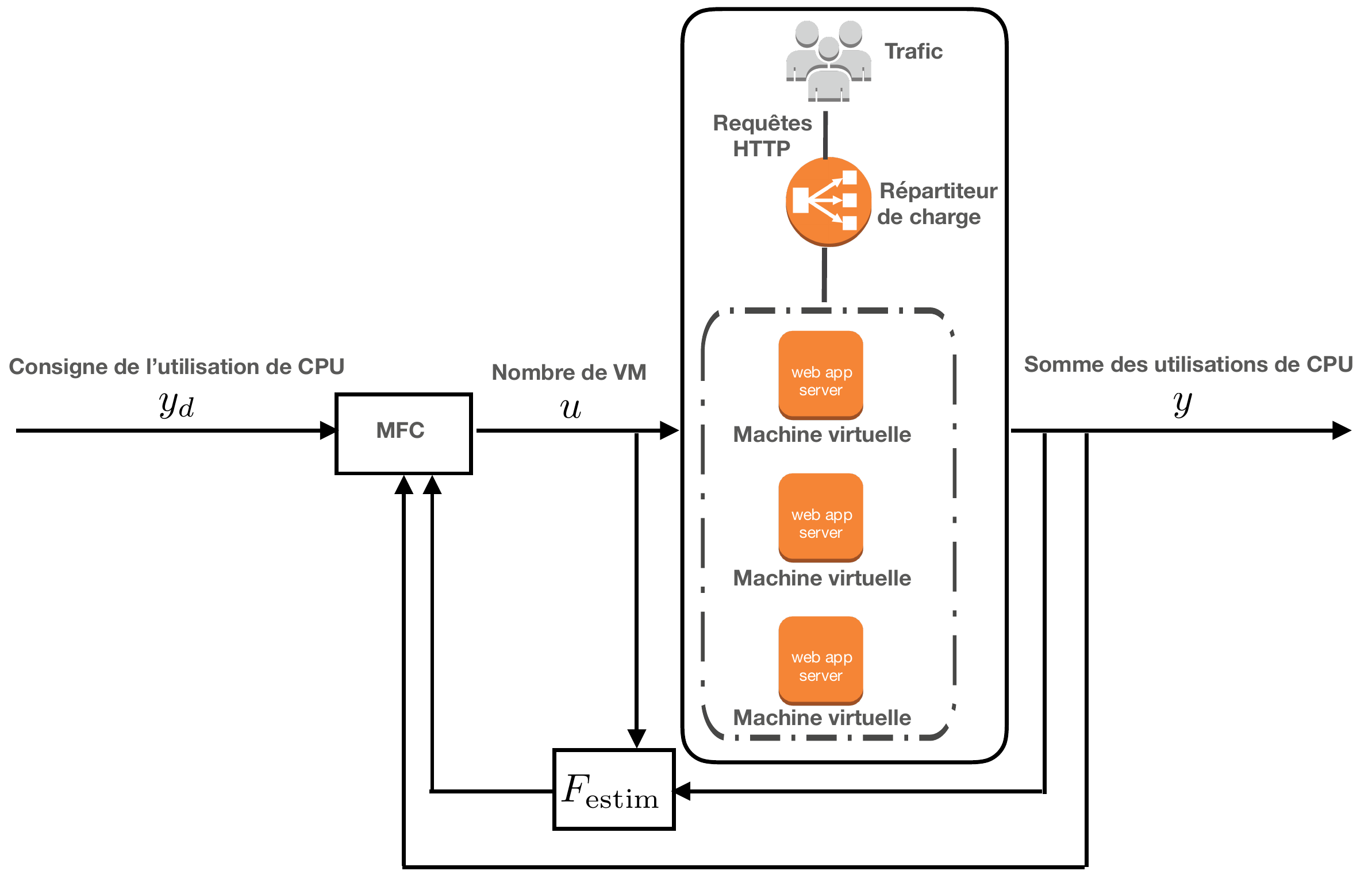}
\caption{Sch\'{e}ma de commande.}\label{Fig_CD}
\end{center}
\end{figure*}

Le sch\'{e}ma bloc {\ref{Fig_CD}} illustre ce qui pr\'{e}c\`{e}de. 
%\newpage

\newpage

\section{Exp\'{e}rimentations}\label{exp}

\subsection{Cadre g\'{e}n\'{e}ral}
\begin{figure*}[h!]
\begin{center}
\includegraphics[width=7in]{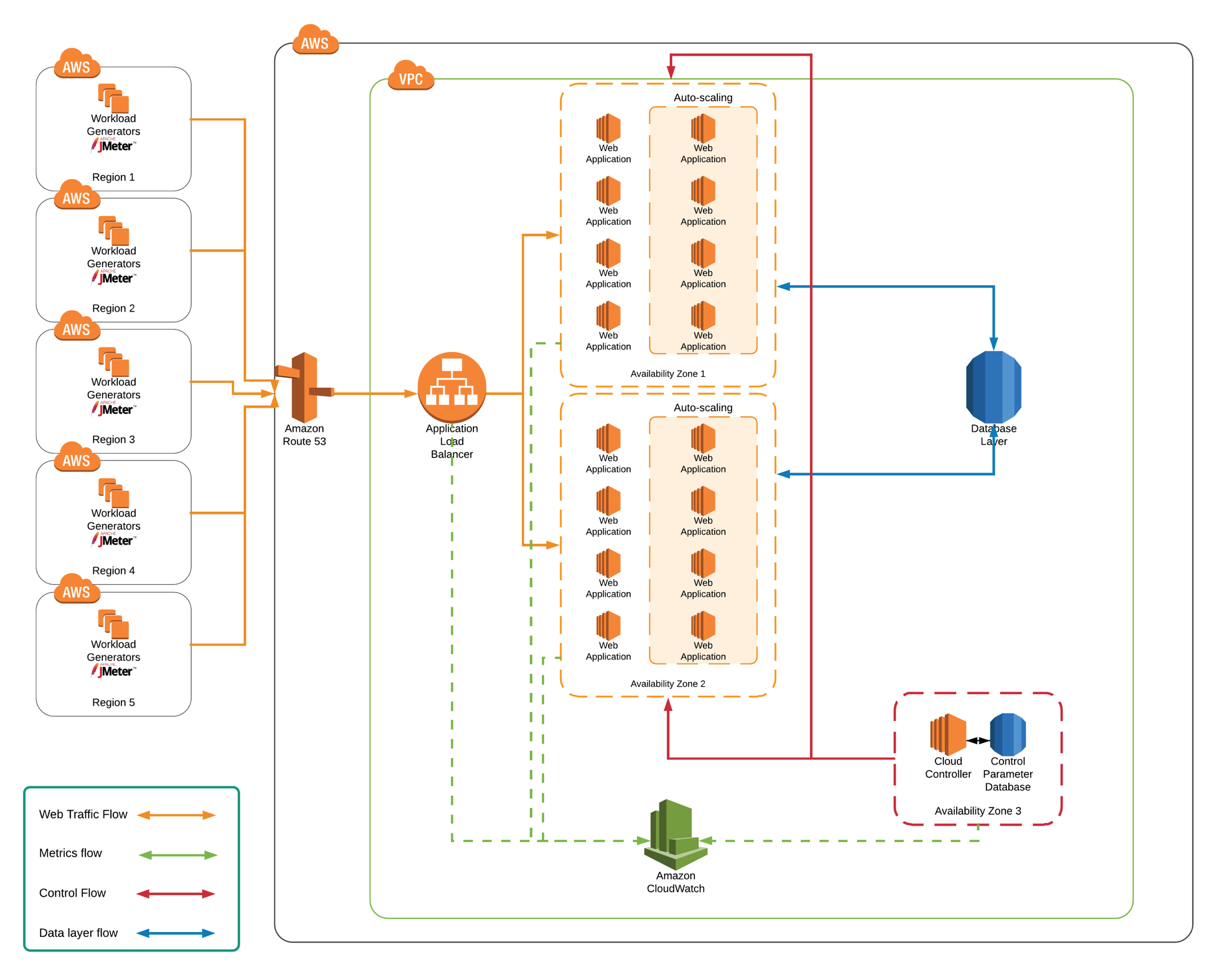}
\caption{Cadre exp\'{e}rimental.}\label{Fig_Frame}
\end{center}
\end{figure*}

La figure  {\ref{Fig_Frame}} repr\'{e}sente l'infrastructure AWS.
\newpage
On stocke, pour se rapprocher autant que faire se peut du fonctionnement authentique, la totalit\'{e} de Wikipedia en anglais (version du 15 novembre 2017). On introduit les deux situations: 
\begin{enumerate}
\item Un trafic en paliers, pour comparer les diverses techniques. 
\item Un trafic  \`a fortes variations, provenant du passage du trafic Wikipedia de 120 \`a 2 heures. En ressort l'excellente r\'{e}activit\'{e} de la commande sans mod\`{e}le. 
\end{enumerate}
On envoie dans les deux cas 1 million de requ\^{e}tes pendant 2 heures.

\subsection{Exp\'{e}riences et comparaisons}\label{pourc}

\begin{figure*}[h!]
\begin{center}
\includegraphics[width=7in]{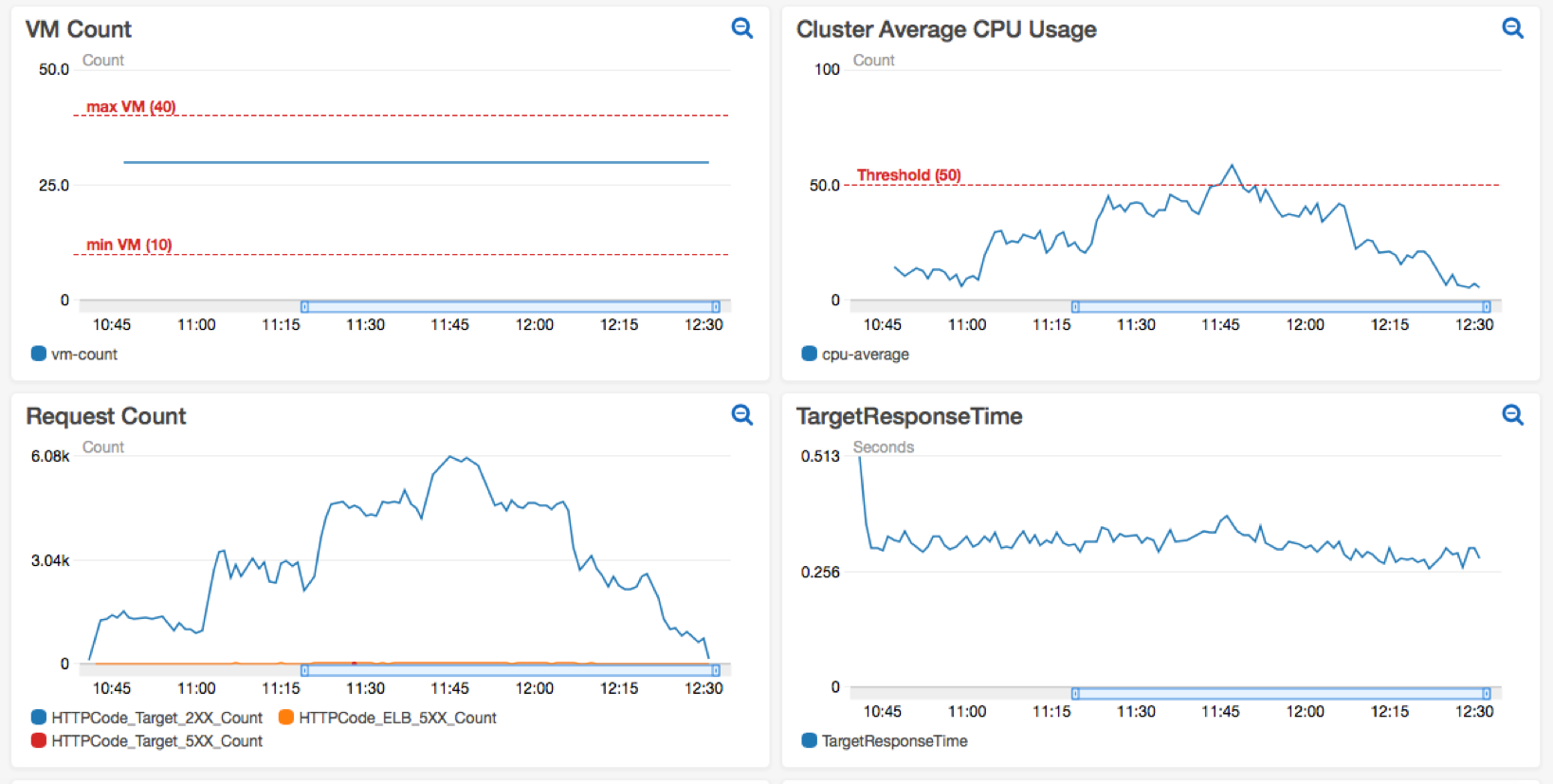}
\caption{R\'{e}sultats exp\'{e}rimentaux sans auto-ajustement, avec $30$ machines virtuelles. }\label{Fig_5}
\end{center}
\end{figure*}

\begin{figure*}[h!]
\begin{center}
\includegraphics[width=7in]{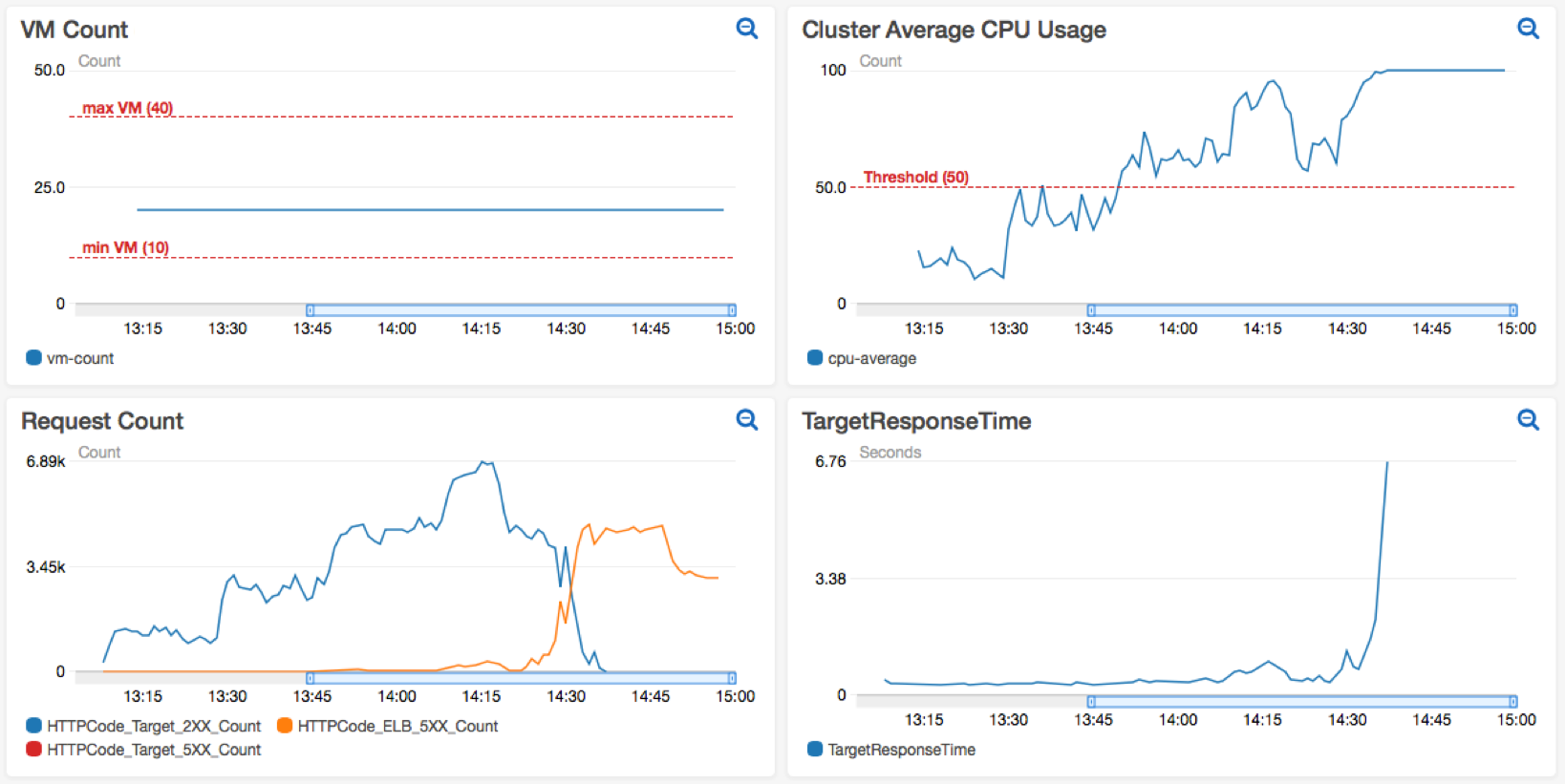}
\caption{R\'{e}sultats exp\'{e}rimentaux sans auto-ajustement, avec $20$ machines virtuelles. }\label{Fig_6}
\end{center}
\end{figure*}

On exp\'{e}rimente d'abord le premier trafic sur une grappe AWS \og{statique}\fg, où le nombre de VM est constant: voir figures {\ref{Fig_5}} et {\ref{Fig_6}}. Quand 
$M_{\rm{act}}^h(t) = 30$, c'est-\`{a}-dire avec 30 VM, pendant le test, la grappe est sur-dimensionn\'{e}e (voir figure {\ref{Fig_5}}). La charge moyenne des processeurs est inférieure à la r\'{e}f\'{e}rence (voir sous-figure \emph{Cluster Average CPU Usage}). 
Avec, en revanche, $M_{\rm{act}}^h(t)=20$ (voir figure {\ref{Fig_6}}), la charge reste au-dessus de la r\'{e}f\'{e}rence pendant les trois quarts de l'exp\'{e}rimentation, et sature. D'où beaucoup d'\'{e}checs, indiqués par la ligne orange de
la sous-figure \emph{Request Count},  et une indisponibilité du service.

On r\'{e}p\`{e}te ce qui pr\'{e}c\`{e}de avec 
\begin{enumerate}
\item l'ajustement AWS pour suivi de cible,
\item  la commande sans mod\`{e}le.
\end{enumerate}

\begin{figure*}[h!]
\begin{center}
\includegraphics[width=7in]{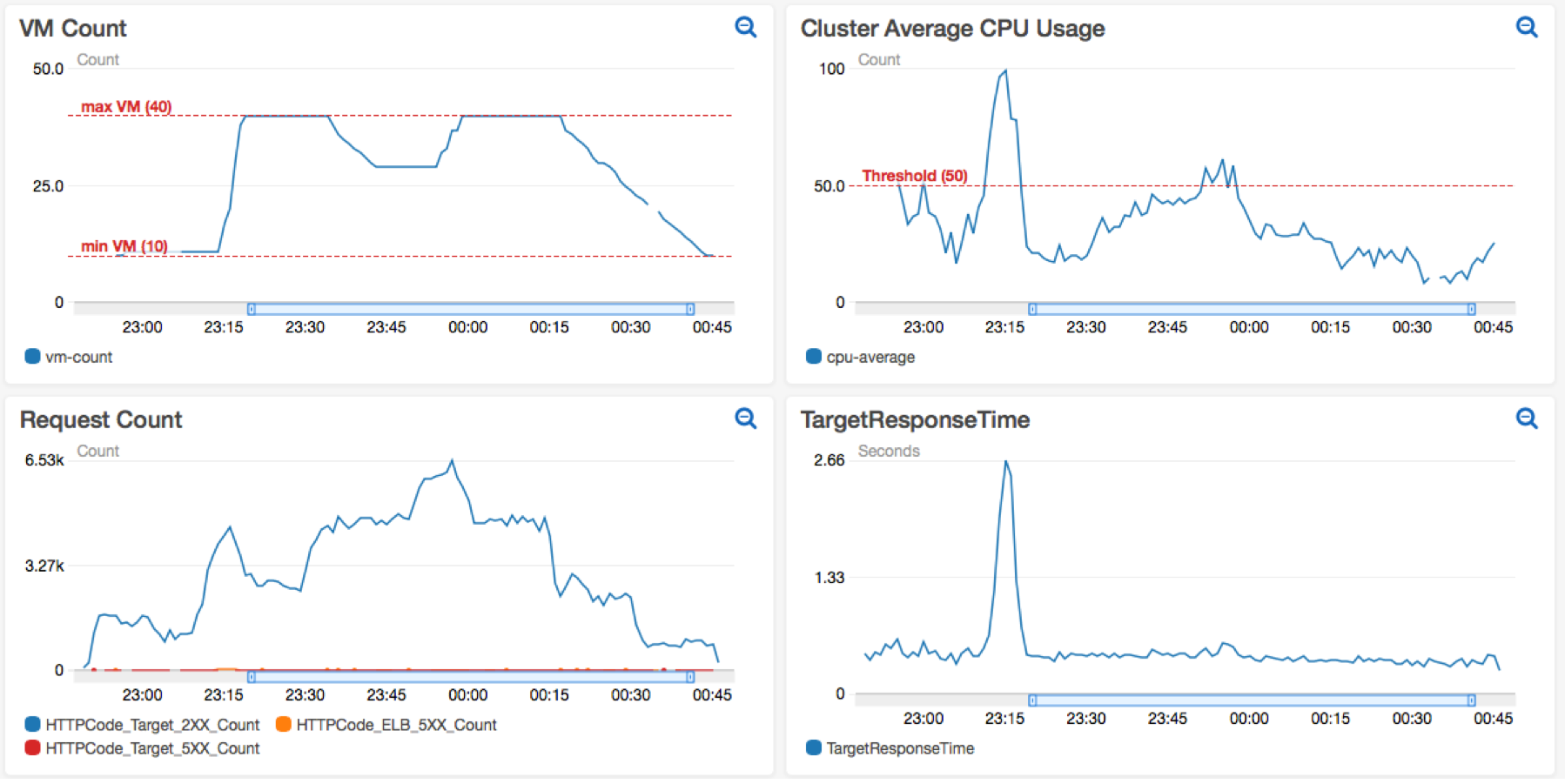}
\caption{R\'{e}sultats exp\'{e}rimentaux avec \textit{AWS Target Tracking Auto-Scaling Algorithm} et variations en \textbf{paliers} du trafic Wikipedia. }\label{Fig_7}
\end{center}
\end{figure*}

La figure {\ref{Fig_7}} décrit les résultats pour la grappe avec l'algorithme \emph{Target Tracking} d'AWS (version d'avril 2018). En fixant la r\'{e}f\'{e}rence à 50\% et les autres param\`{e}tres à leurs valeurs proposées par Amazon. Cette procédure d\'{e}tecte le premier pic de trafic. Le nombre de VM croît alors jusqu'au maximum autoris\'{e}. Il diminue apr\`{e}s d\'{e}tection d'une sous-utilisation. D'o\`{u} un retard. Soulignons les faits suivants: 
\begin{itemize}
\item l'utilisateur ne peut changer la fréquence d'échantillonnage et les seuils d'activation,
\item Amazon ne d\'{e}taille pas la d\'{e}termination du nombre de VM.
\end{itemize}

\begin{figure*}[h!]
\begin{center}
\includegraphics[width=7in]{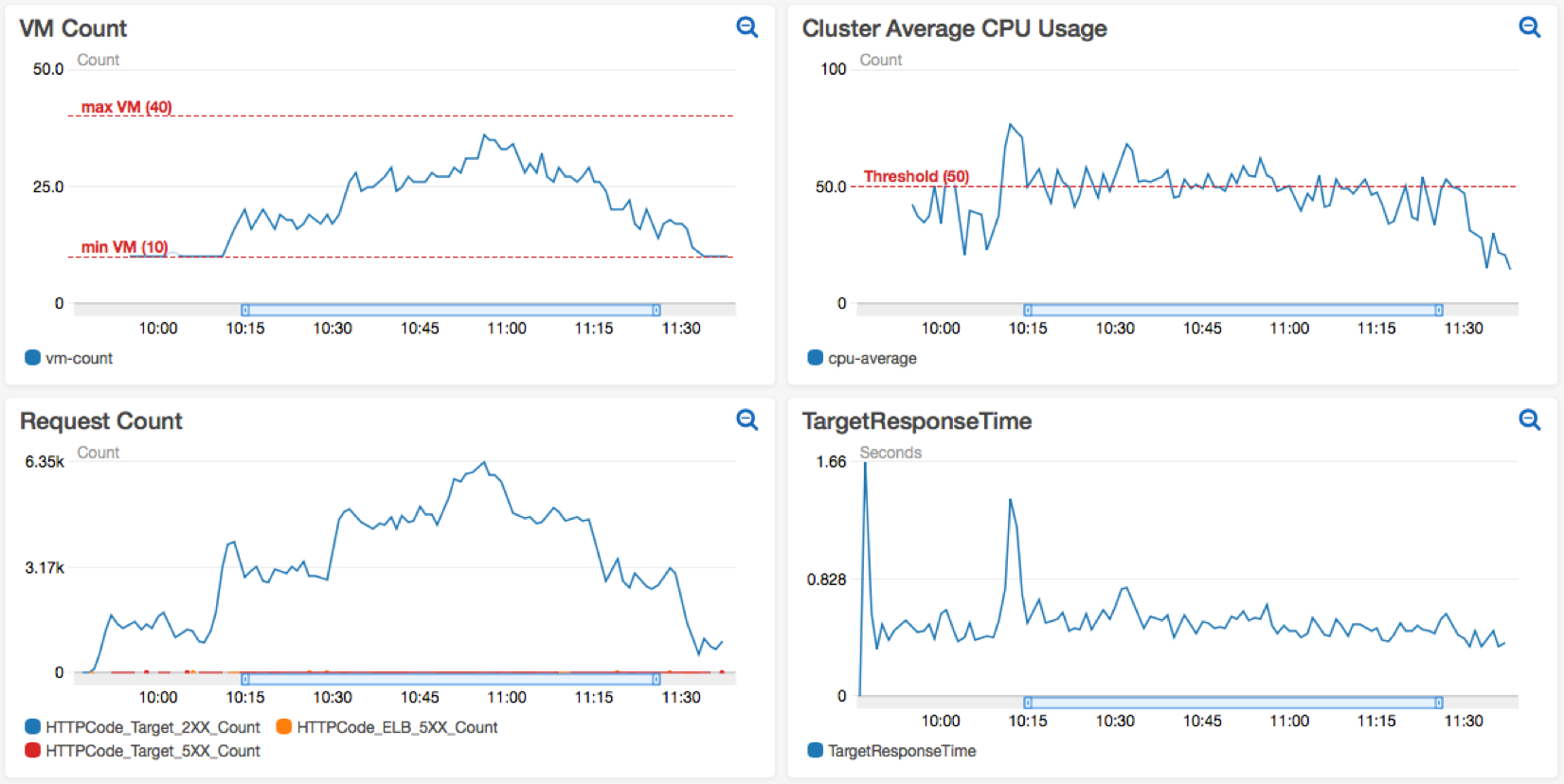}
\caption{R\'{e}sultats exp\'{e}rimentaux avec commande sans modèle et variations en \textbf{paliers} du trafic Wikipedia.} \label{Fig_1}
\end{center}
\end{figure*}

\begin{figure*}[h!]
\begin{center}
\includegraphics[width=7in]{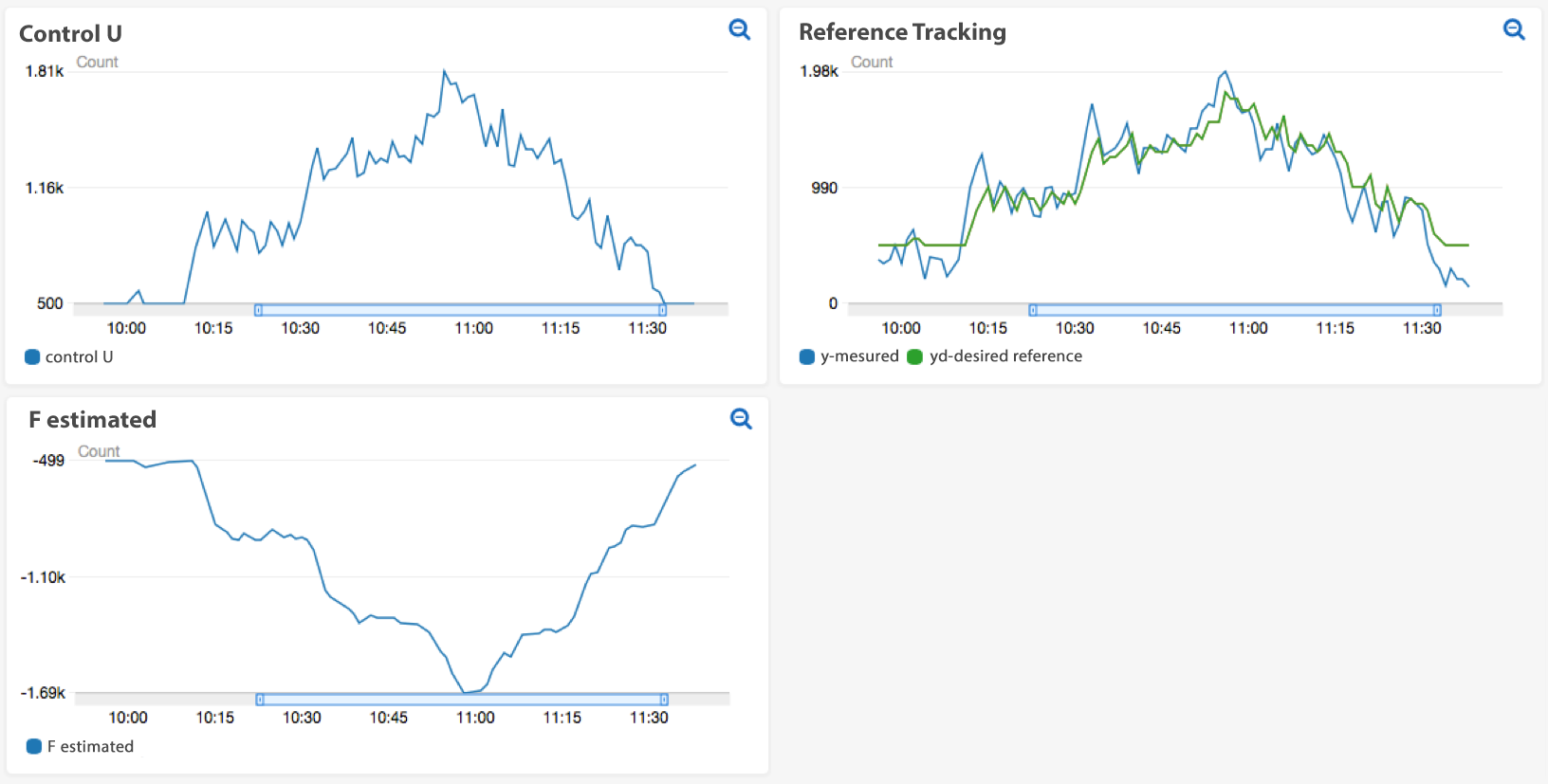}
\caption{R\'{e}sultats avec correcteur iP exp\'{e}rimental ($K_p = 0.8$, $\alpha = 1$, échantillonnage: 1 min) et variations en \textbf{paliers} du trafic Wikipedia.}\label{Fig_2}
\end{center}
\end{figure*}

Passons à la commande sans modèle. On observe en figure {\ref{Fig_1}} l'adaptation parfaite de la grappe $M_{\rm{act}}^h(t)$ (sous-figure \emph{VM count}) au nombre 
de requ\^{e}tes. La sous-figure \emph{TargetResponseTime} décrit un temps de r\'{e}ponse à peu près constant, sauf lors de la premi\`{e}re apparition d'un pic de charge. 
En figure {\ref{Fig_2}}, sous-figure \emph{Reference Tracking}, on observe un suivi impeccable de l'objectif d\'{e}sir\'{e}. La courbe bleue représente la somme des utilisations des processeurs mesurées sur chaque VM (\textit{cf.} \eqref{Ymesured}) et la courbe verte représente la consigne désirée (\textit{cf.} \eqref{Ydesired}).
 La sous-figure \emph{F estimated} dépeint une excellente estimation
de l'allure de la charge.

\begin{figure*}[h!]
\begin{center}
\includegraphics[width=7in]{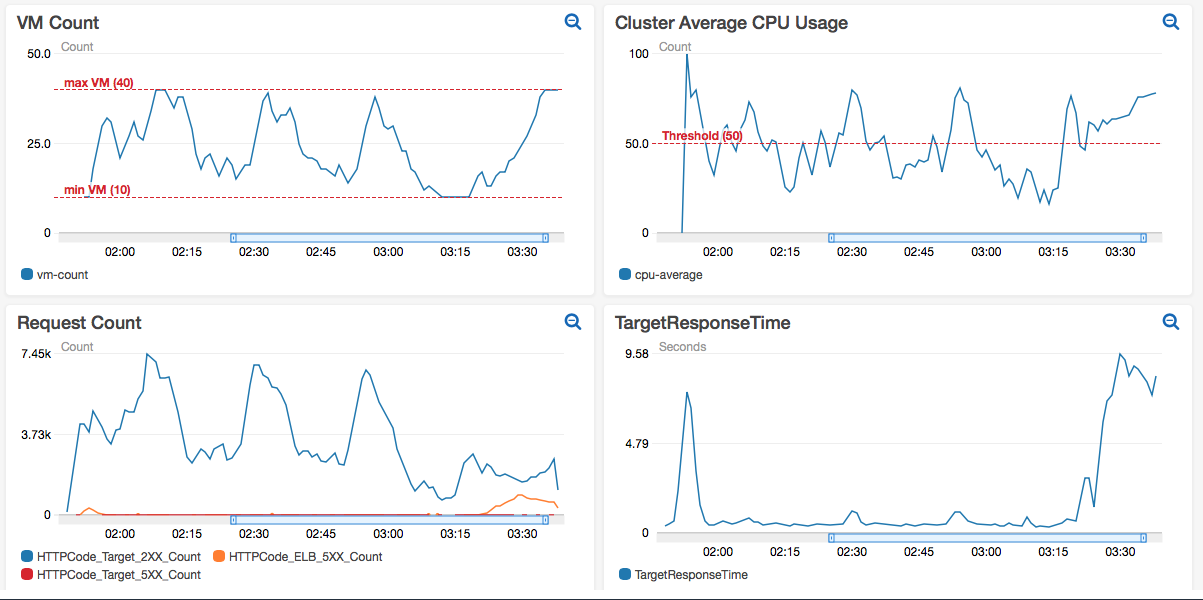}
\caption{R\'{e}sultats exp\'{e}rimentaux avec commande sans modèle et variations \textbf{aig\"{u}es} de trafic
Wikipedia.}\label{Fig_3}
\end{center}
\end{figure*}

\begin{figure*}[h!]
\begin{center}
\includegraphics[width=7in]{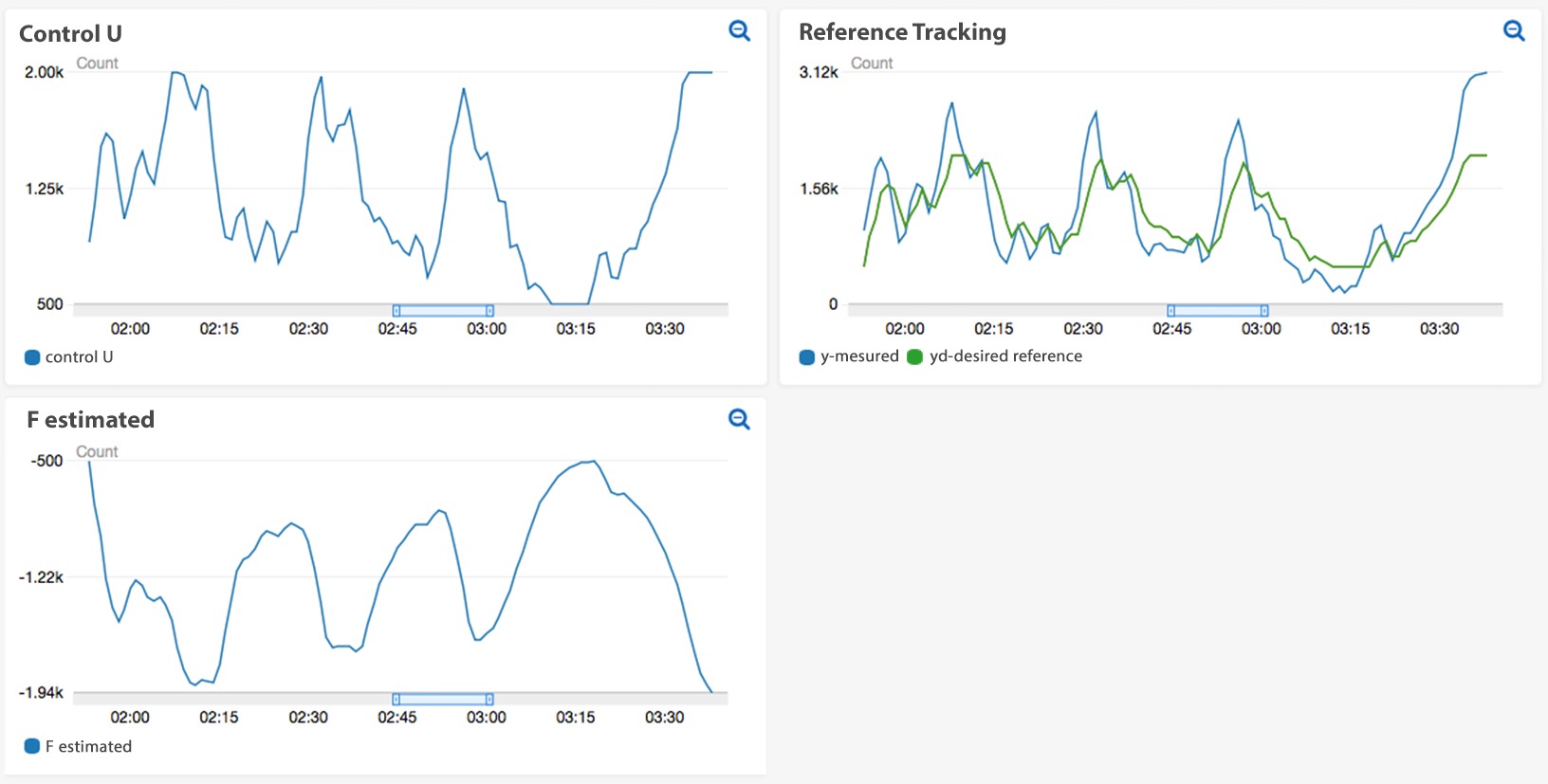}
\caption{R\'{e}sultats avec correcteur iP exp\'{e}rimental ($K_p = 0.8$, $\alpha = 1$, échantillonnage: 1 min) et variations \textbf{aig\"{u}es} du trafic Wikipedia pour la grappe AWS.}\label{Fig_4}
\end{center}
\end{figure*}

Le r\'{e}partiteur de charges Amazon \emph{Elastic Load Balancing} (\emph{ELB}), du type \emph{Application}\footnote{Voir \\
  {\tt https$:$//docs.aws.amazon.com/fr\_fr/elasticloadbalancing/latest/ \\application/introduction.html}}, est conçu pour s'adapter à l'arrivée des requ\^{e}tes. Il pr\'{e}sente toutefois une limitation: pendant un laps de temps de 5 minutes, le trafic ne doit pas croître de plus de 50\% \footnote{Voir \\
  {\tt https$:$//aws.amazon.com/articles/best-practices-in-evaluating-elastic-load-balancing}}. C'est pourquoi on introduit, ici, des variations très brutales, peu communes en pratique, pour affiner la confrontation avec le sans-modèle. D'après les figures {\ref{Fig_3}} et {\ref{Fig_4}}, l'iP \eqref{ip} suit l'allure de la charge et adapte au mieux la taille de la grappe.

\newpage

On construit le tableau {\ref{Tab_1}} grâce aux \og{indicateurs de performance}\fg, ou  \emph{Key Performance Indicators} (\emph{KPI}), suivants:
\begin{itemize}
\item somme des durées de vie, en secondes, des VM utilis\'{e}es pendant une exp\'{e}rimentation,
\item d\'{e}viation de mesure du capteur, c'est-à-dire de la moyenne de charge du processeur, par rapport \`a la r\'{e}f\'{e}rence 50\% (voir les sous-figures Cluster Average CPU Usage).
\end{itemize}
Ils précisent les termes de la comparaison: l'auto-ajustement par commande sans mod\`{e}le induit une forte diminution de la consommation temporelle des VM, donc du co\^{u}t, tout en assurant un suivi excellent de la r\'{e}f\'{e}rence.

 \begin{table} 
\begin{center}
\begin{tabular}{| l | p{6cm} | p{5.8cm} |}
  \hline
\textbf{Technique} & \textbf{Somme des durées de vie, en secondes, des VM utilis\'{e}es}&\textbf{D\'{e}viation moyenne du CPU par rapport \`a  la référence}\\
  \hline
  Commande sans mod\`{e}le (fig. {\ref{Fig_1}}) & 127 920 & 8,53\%\\
AWS Target Tracking (fig. {\ref{Fig_7}})& 187 080  & 21,73\% \\
Sans \'{e}lastisit\'{e}, avec 20 VM (fig. {\ref{Fig_6}}) & 144 000  & 28,36\%\\
Sans \'{e}lastisit\'{e}, avec 30 VM (fig. {\ref{Fig_5}})& 216 000 & 21,78\% \\
  \hline
\end{tabular}
\end{center}
\caption{Comparaison des techniques.} \label{Tab_1}
\end{table}

\begin{remark}
Les comparaisons faites avec d'autres techniques, comme les PID, semblent tout aussi favorables \`{a} notre approche. L'absence de d\'{e}tails dans les publications, que nous avons pu lire, nous interdit ici d'en dire plus.
\end{remark}

\newpage

\pagebreak

\section{Conclusion}\label{conclusion}
La \og{tolérance aux pannes}\fg, ou \emph{fault tolerance}, sujet classique en automatique (voir, par exemple, \cite{blanke}), appara\^{\i}t évidemment en nuagique (voir, par exemple, \newline \cite{arab}). La défaillance de machines virtuelles y est un enjeu saillant. Comme la cardinalité des VM est la commande, on sait \cite{ijc13,toulon} que la commande sans modèle surmonte aisément ces aléas. De futures publications le confirmeront.

De prochains travaux \'{e}tendront notre m\'{e}thode aux \og conteneurs \fg, ou \emph{containers}, (voir, par exemple, \cite{bernstein,pahl}) d\'{e}velopp\'{e}s notamment par la compagnie Docker, de plus en plus employ\'{e}s. Signalons, par exemple, \cite{baresi} pour une \'{e}lasticit\'{e} par r\'egulation \`a partir d'un mod\`ele simplifi\'e.  Bien d'autres th\`{e}mes, assez voisins de ce travail, m\'{e}ritent \'{e}galement consid\'{e}ration.
Mentionnons-en un. Une d\'{e}marche analogue devrait permettre d'aborder les n{\oe}uds de calcul, 
comme \emph{MapReduce}, d\^{u} \`a Google. On les utilise pour l'analyse des 
\og{m\'{e}gadonn\'{e}es}\fg, ou \emph{big data}. Les requ\^{e}tes pr\'{e}c\'{e}dentes deviennent 
celles des d\'{e}cideurs \`a propos des \og{jeux de donn\'{e}es}\fg, ou \emph{datasets} \cite{Berekmeri2016,Cerf2016CDC,cerf}. 

Si \og{les techniques sont des proc\'{e}d\'{e}s bien d\'{e}finis et transmissibles destin\'{e}s 
\`a produire certains r\'{e}sultats jug\'{e}s utiles}\fg \ \cite{lalande}, le nuagique est, 
alors, une collection de techniques plutôt qu'une science. 
Le contenu de cet article devrait, s'il se confirme, infl\'{e}chir cet \'{e}tat, du moins partiellement. Ce serait, aussi, le signal, longtemps pressenti, de la place \'{e}minente qui 
revient \`a l'automatique en informatique, surtout si l'on abandonne l'ambition, trop souvent 
vaine, pour ne pas dire na\"{\i}ve, d'une mod\'{e}lisation math\'{e}matique pr\'{e}cise, qu'elle 
soit d\'{e}terministe ou probabiliste. Que les math\'{e}maticiens  se rassurent! Leur r\^{o}le 
ne sera en rien diminu\'{e} s'ils participent \`a l'\'{e}dification des nouveaux outils 
exig\'{e}s, dont la commande sans mod\`{e}le n'est qu'un exemple parmi bien d'autres \`a d\'{e}velopper\footnote{Voir, par exemple, \cite{solar} sur les \og{chroniques}\fg, ou \emph{time series}: \og{m\'{e}gadonn\'{e}es}\fg, ou \emph{big data}, et \og{apprentissage}\fg, ou \emph{machine learning}, y jouent un r\^{o}le moindre qu'ailleurs. Il convient de mentionner que des points de vue classiques sur les chroniques ont d\'{e}j\`a \'{e}t\'{e} employ\'{e}s assez souvent en nuagique (voir, par exemple, \cite{alk,cal} et leur bibliographie).} sinon \`a inventer.

%D'autres applications nuagiques sont \`a  explorer, comme les conteneurs et leur orchestration. 
%Devenus \`a  la mode depuis plusieurs ann\'{e}es, les conteneurs comme Docker sont largement utilis\'{e}s 
%par de nombreuses entreprises comme Google, Microsoft, IBM, etc. Afin de pouvoir b\'{e}n\'{e}ficier 
%\xout{\info{au maximum de cet approche et}} de la \xout{\info{fine}} granularit\'{e} et 
%\info{de l'}agilit\'{e} 
%qu'offrent les conteneurs, des
%orchestrateurs ont vu le jour comme Kubernetes (d\'{e}velopp\'{e} par Google) ou OpenShift (d\'{e}velopp\'{e} 
%par RedHat). Ces orchestrateurs impl\'{e}mentent des m\'{e}canismes d'auto-ajustement bas\'{e}s sur 
%\info{des} 
%algorithmes simples de d\'{e}passement de seuil. Une application de l'algorithme d'auto-ajustement 
%de la Commande Sans Mod\`{e}le nuagique semble pertinente et sera pr\'{e}sent\'{e}e lors de nos futures 
%publications.

%\section{ToBeCited}
%
%  \cite{Verma2011};
% Survey:
% \cite{Zhang2010};  \\
% Control-riented: Fuzzy \cite{Grimaldi2017} ; \cite{Fokaefs2017} ; \cite{Fokaefs2016} ;  \\
% 
% Autoscaling: \cite{Qu2017} ; \cite{Singh2015} \cite{Galante2012}; \cite{Lim2010} ; linear parameter- varying (LPV) systems (private cloud), \cite{Saikrishna2017}

\end{document}